\documentclass[aps,prl,twocolumn,showpacs,superscriptaddress,groupedaddress,longbibliography]{revtex4-1}
\usepackage{graphicx}
\usepackage{color}
\usepackage{here}

\def \beq{\begin{equation}}
\def \eeq{\end{equation}}
\def\eqref#1{(\ref{#1})}
\def\bea{\begin{eqnarray}}
\def\eea{\end{eqnarray}}
\def\jpsi{J\kern-0.1em/\kern-0.1em\psi}

\def\URLtilde{\lower0.2em\hbox{$\tilde{\phantom{a}}$}}
\def\mycomm#1{\hfill\break\strut\kern-3em{\color{red}\tt ====> #1
\color{black}}\hfill\break}

\begin{document}

\setcounter{footnote}{1}
% \thispagestyle{empty}
%\rightline{EFI 21-XX}
%\rightline{TAUP YYYY/21}
%\rightline{arXiv:2111.nnnnn}
%\vskip1.5cm

%\begin{center}
%{\large \bf \boldmath
	\title{The doubly charmed strange tetraquark} 
%\unboldmath}
%\end{center}
%\bigskip

\author{Marek Karliner}
\email{marek@tauex.tau.ac.il}
\affiliation{School of Physics and Astronomy \\
%Raymond and Beverly Sackler Faculty of Exact Sciences \\
Tel Aviv University, Tel Aviv 69978, Israel}
\author{Jonathan L. Rosner}
\email{rosner@hep.uchicago.edu}
\affiliation{Enrico Fermi Institute and Department of Physics \\
University of Chicago, 5640 S. Ellis Avenue, Chicago, IL
60637, USA}

\date{January 5, 2022}
\begin{abstract}
The LHCb experiment at CERN has discovered a doubly charmed isoscalar
tetraquark $T_{cc}$ with the quantum numbers of $c c \bar u \bar d$ and mass
of about 3875 MeV/$c^2$, decaying to $D^0 D^0 \pi^+$ through the intermediate
channel $D^{*+} D^0$.  We present a study of its strange companions with the
quantum numbers of $c c \bar q \bar s$, where $q = u, d$ and isospin violation
is neglected.
\end{abstract}
\maketitle

\sloppy
The LHCb Experiment at CERN has discovered an exotic meson $T_{cc} = c c \bar u
\bar d$ \,decaying to $D^0 D^0 \pi^+$ through the channel $D^{*+} D^0$, with a
mass around 3875 MeV/$c^2$ \cite{LHCb:2021vvq,LHCb:2021auc}.  This mass is just
below that for which the pion can form a $D^{*+}$ resonance with either $D^0$.
It is also within 7 MeV/$c^2$ of the value of $3882 \pm 12$ MeV/$c^2$
predicted in Ref.\ \cite{Karliner:2017qjm} using the same quark-model
parameters that led to the successful prediction \cite{Karliner:2014gca} of
the mass of the doubly charmed baryon $\Xi_{cc}^{++}$ \cite{LHCb:2017iph}.
We use the methods of Ref.\ \cite{Karliner:2017qjm} to predict the masses of
the strange companions of the $T_{cc}$, and propose a study of the $D^0 D^0
K^+$ system to search for excited companions of the doubly charmed strange
tetraquark.  We comment on production and decay of the $T_{cc,s}$ states.

A generalization of Table II
in Ref.\ \cite{Karliner:2017qjm} gives a spin-1 state at 4106 MeV which can
decay to $D^* D_s$ or $D D_s^*$, composed of a $c c$ diquark with spin 1 and a
$\bar q \bar s~(q=u,d)$ antidiquark with spin zero.  We reproduce the table in
Ref.\ \cite{Karliner:2017qjm} which led to the successful prediction of the
$T_{cc}$ mass (Table \ref{tab:ccud}) and then indicate the changes that
apply to a $T_{cc;s} = c c \bar q \bar s$ (Table \ref{tab:ccus}).  Here we have
taken $m^b_q = 363$ MeV, $m^b_s = 538$ MeV, and the superscript $b$ denotes
values suitable for baryons (and tetraquarks).

There are also states with spin 0, 1, and 2 composed of a $c c$ diquark with
spin 1 and a $\bar q \bar s$ with spin 1.  The spin-weighted average mass
$\bar M$ of the corresponding hyperfine multiplet differs from 4106 MeV
only in the mass of the light $\bar q \bar s$ diquark.  For a nonstrange
diquark this ``bad"$-$``good" difference has long been known 
to be 200 MeV=$(2/3)[M(\Delta)-M(N)]$, as in \cite{Karliner:2014gca} and earlier references therein.
It is encouraging that a recent lattice QCD calculation \cite{Francis:2021vrr}
finds this difference to be 198(4) MeV.  The corresponding difference for a
strange diquark is $(m^b_q/m^b_s)\cdot 200$ MeV = 135 MeV,
so the spin-weighted average mass 
of the hyperfine multiplet with two spin-1 $cc$ and $\bar q \bar s$ diquarks
is obtained by adding the ``bad"$-$``good" 
$\bar q \bar s$ diquark mass difference to the state in Table II,
$\bar M = 4106 + 135 = 4241$ MeV. 

The two spin-1 $cc$ and $\bar q \bar s$ diquarks can form $S$-wave states
with with spin $(0, 1, 2)$. The mass splittings between the members of 
this multiplet result only from hyperfine, i.e., spin-dependent interaction 
between the two diquarks. They scale like $\Delta  M = (-2x, -x, x)$ for
spin $(0, 1, 2)$, respectively.
The $D^0 D^0 K^+$ threshold is at 4223 MeV.  Depending on
the sign of $x$, at least either the spin-zero or spin-2 three-body
resonance should be able to decay to $D^0 D^0 K^+$.
Indeed, we expect the spin-spin interaction to be repulsive, i.e< $x > 0 $.
In quark-model calculations, whether it is quark-quark or
quark-antiquark, we always have that spin-spin interaction is
antiferromagnetic, i.e., spin 1 has higher energy than spin zero.
This is likely to carry over into interaction
of two spin-1 diquarks, since such an interaction presumably can 
be viewed as the sum of interactions of quarks in one diquark with 
 quarks in the other diquark.

% This is Table I
\begin{table}%[H]
\caption{Contributions to the mass of the lightest tetraquark
$T(cc\bar u\bar d)$ with two charmed quarks and $J^P=1^+$.
\label{tab:ccud}}
\begin{center}
\begin{tabular}{c r} \hline \hline
Contribution & Value (MeV) \\ \hline
$2m^b_c$ & 3421.0\\
$2m^b_q$ & 726.0\\
$a_{cc}/(m^b_c)^2$ & 14.2 \\
${-}3a/(m^b_q)^2$ & ${-}150.0$ \\
$cc$ binding & ${-}129.0$ \\
Total & $3882.2 \pm 12$ \kern-2.55em \\ \hline \hline
\end{tabular}
\end{center}
\end{table}

% This is Table II
\begin{table}%[H]
\caption{Contributions to the mass of the lightest tetraquark $T(cc\bar q
\bar s)$ with two charmed quarks, $\bar q \bar s$ in a state of spin
zero, and $J^P=1^+$.
\label{tab:ccus}}
\begin{center}
\begin{tabular}{c r} \hline \hline
Contribution & Value (MeV) \\ \hline
$2m^b_c$ & 3421.0\\
$m^b_q + m_s$ & 901.0\\
$a_{cc}/(m^b_c)^2$ & 14.2 \\
${-}3a/(m^b_q m^b_s)$ & ${-}101.2$ \\
$cc$ binding & ${-}129.0$ \\
Total & $4106 \pm 12$ \kern-2.55em \\ \hline \hline
\end{tabular}
\end{center}
\end{table}

The identification of states decaying to $D_s$ or $D_s^*$ poses significant
challenges.  The largest exclusive branching fraction of the $D_s$ is
to $K^+ K^- \pi^+$, with ${\cal B} = (5.39 \pm 0.15)\%$
\cite{ParticleDataGroup:2020ssz}. Furthermore, the
soft photon in $D_s^* \to \gamma D_s$ will be very difficult to identify,
   preventing full reconstruction of the $D_s^*$ decay.  Table \ref{tab:recon}
lists some prospective final states of the predicted $T_{cc,s}(4106)$.  The
most promising decay is $T_{cc,s}^{++} \to D^{*+} D_s$, where $D^{*+} \to
D^0 \pi^+$ (giving an identifiable soft pion) and $D_s \to K^+ K^- \pi^+$
(giving a fully reconstructed final state).

% This is Table III
\begin{table}
\caption{ Decay modes of the ground-state singly and doubly charged $T_{cc,s}$
with spin 1 and mass 4106 MeV.
\label{tab:recon}}
\begin{center}
\begin{tabular}{ccc} \hline \hline
     Decaying    &        Quark         & Prospective \\
       state     &       content        & final state \\ \hline
  $T_{cc,s}^+$   & $cc {\bar u \bar s}$ & $D^0 D_s^{*+}$ \\
        	 &                      & $D^{*0}D_s^+$ \\
 $T_{cc,s}^{++}$ & $cc {\bar d \bar s}$ & $D^+ D_s^{*+}$ \\
                 &                      & $D^{*+} D_s^+$ \\
\hline \hline
\end{tabular}
\end{center}
\end{table}

Other calculations of $M(T_{cc,s})$ include, e.g., 3975, 3979 MeV for the singly
and doubly charged state \cite{ren:2021dsi} based on a molecular picture, and
4156 MeV \cite{Eichten:2017ffp}, based on heavy-quark symmetry (giving 3978 MeV
for the nonstrange state). A comprehensive list of theoretical mass predictions
for the $T_{cc}$ states can be found in Refs.~\cite{LHCb:2021vvq,LHCb:2021auc}.

The predictions for $T_{cc}$ and $T_{cc,s}$ masses in Tables I and II are 
based on the same approach. Therefore, if it turns out that the mass of the
lightest doubly-charmed strange tetraquark $M(T_{cc,s})$ is significantly
different from 4106 MeV, it will imply that LHCb's $T_{cc}$ candidate reported
in \cite{LHCb:2021vvq,LHCb:2021auc} 
is unlikely to be the state predicted in Ref.\ \cite{Karliner:2017qjm}. 
If so, the most probable interpretation will be a molecular state, but
one also needs to examine the possibility that it is a kinematic effect, as
discussed below.

\begin{figure}[t]
\begin{center}
\includegraphics[width=0.48\textwidth]{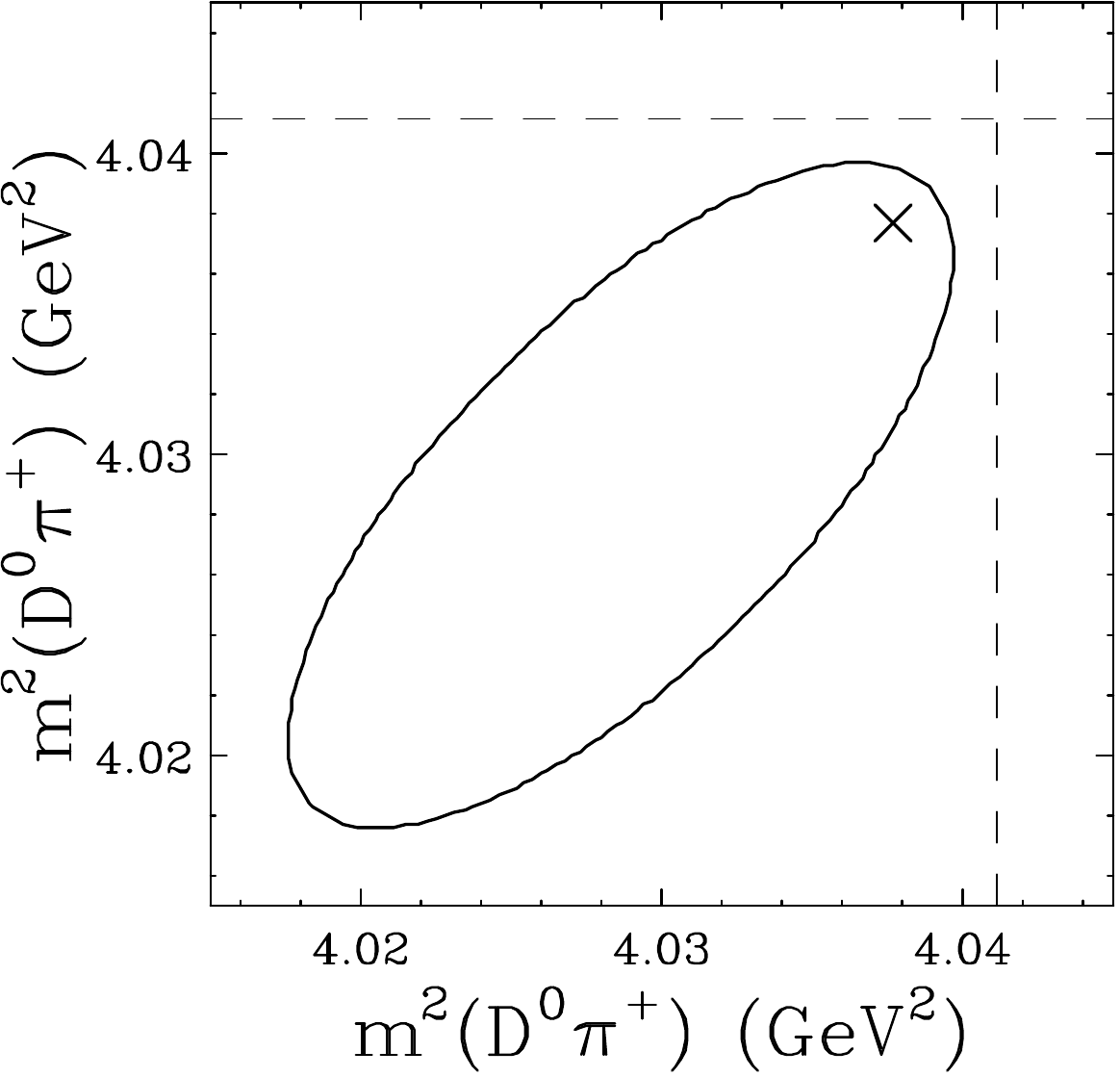}
\end{center}
\caption{Solid curve:
boundary of $D^0 D^0 \pi^+$ Dalitz plot for a center-of-mass (c.m.)
 energy of $M_0 = 3874.7$ MeV.  Dashed straight lines denote central values of
 $m(D^{*+}) = 2010.26$ MeV.  The cross marks the approximate maximum of the
two-dimensional distribution.  Axis units are in GeV$^2$.
}
\label{fig:ddpi}
\end{figure}

The LHCb analysis of the
$D^0 D^0 \pi^+$ system via a unitarized Breit-Wigner formalism gives rise to a
resonance at a mass of $361 \pm 40$ keV below $D^{*+} D^0$ threshold, or at
approximately $M_0=3874.7$ MeV.  (We shall use units in which $c = \hbar = 1$.)
We show the boundary of the $D^0D^0\pi^+$ Dalitz plot along with the
maximum of the two-dimensional distribution in Fig.\ \ref{fig:ddpi}.  The
proximity of this maximum to the intersection of the two $M(D^{*+})$ dashed
straight lines is a cautionary signal of a possible kinematic enhancement.

%\onecolumngrid
% This is Figure 1

The lowest-lying
$D^0 K^+$ resonant subsystem in the three-body $D^0 D^0 K^+$ system is called
$D_{s1}^{*+}(2700)$ in Ref.\ \cite{ParticleDataGroup:2020ssz}. Its mass is
$2714 \pm 5$ MeV and its width is $122 \pm 10$ MeV.  Henceforth we shall refer
to this resonance as $D_s(2714)$.  With its spin-parity $1^-$ and
its mass about 600 MeV above the $D^*_s(2112)$ it is a candidate for a
2S radial excitation of that state.  

% This is Figure 2
\begin{figure}
\begin{center}
\includegraphics[width = 0.48\textwidth]{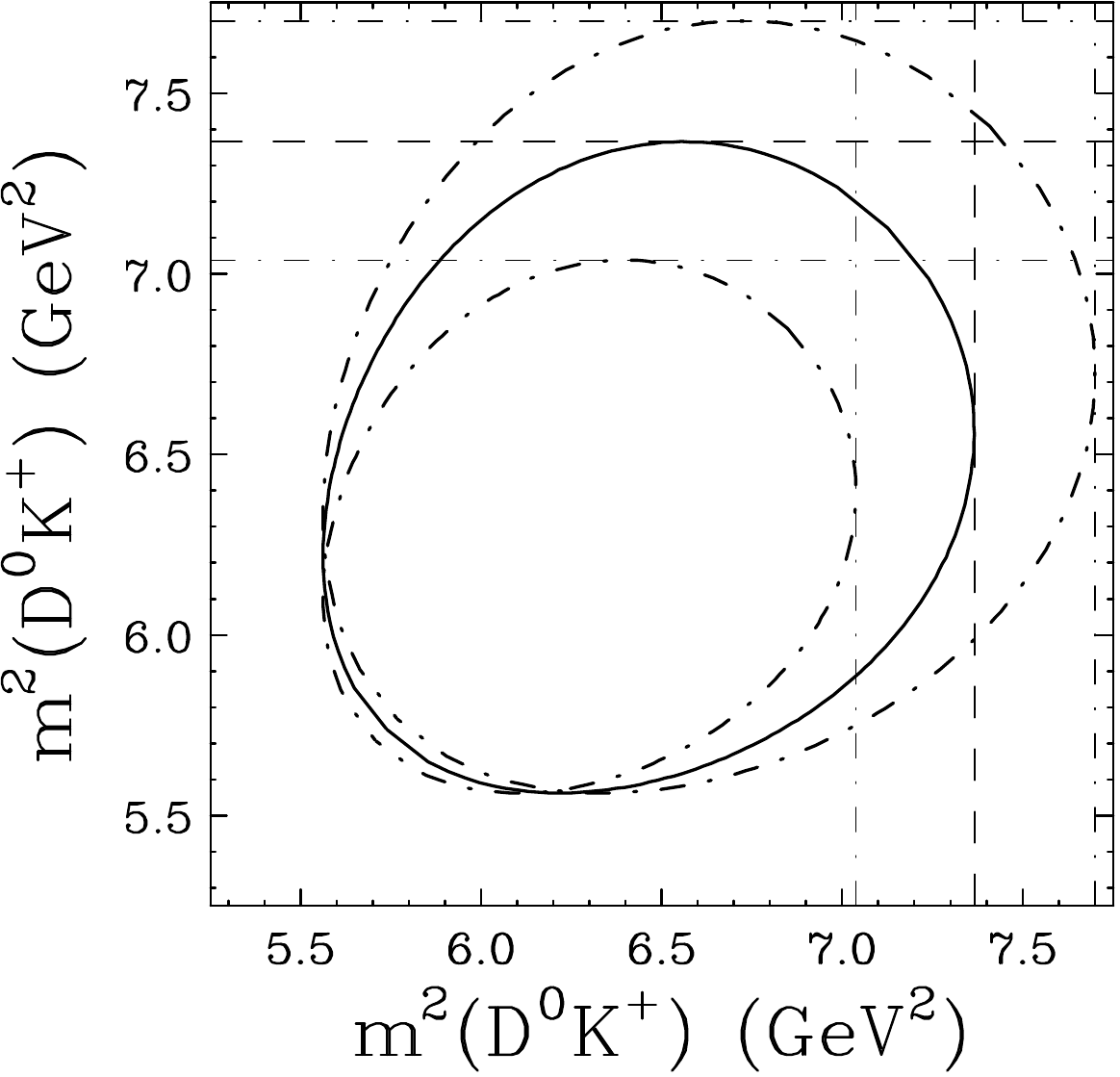}
\end{center}
\caption{\vrule width 0pt height 3ex % to fine-tune vertical placement
Boundary of $D^0 D^0 K^+$ Dalitz plot for center-of-mass (c.m.) energy
of 4588 MeV.  Axes denote squared effective masses of the two $D^0 K^+$
 combinations.  Dashed straight lines correspond to $m(D^0 K^+) =  2714$ MeV.
Dot-dashed ovals correspond to c.\ m.\ energy 4527 and 4649 MeV.  Dot-dashed
straight lines correspond to $M(D_{s1}) = 2714 \pm 61$ MeV.
\label{fig:ddk2}}
\end{figure}
\twocolumngrid

The boundary of the Dalitz plot in Fig.~2 is for a value of $M(D^0 D^0 K^+) =
4588$ MeV which makes it just tangent to the $D^0 K^+$ resonance band at 2714
 MeV.  One is then invited to look for a peak near 4588 MeV in the distribution
of $M(D^0 D^0 K^+)$.  If one is seen, it could indicate that the tangency
condition helps to generate a three-body resonance with quark content $c c
\bar q \bar s$, though probably a broad one in view of the large width of
the $D_{s1}(2714)$.  The dot-dashed ovals and straight lines correspond to
displacing $M(D_{s1}$ by $\pm\Gamma/2$ from its central value.

% This is Figure 3
\begin{figure}
\begin{center}
\includegraphics[width=0.48\textwidth]{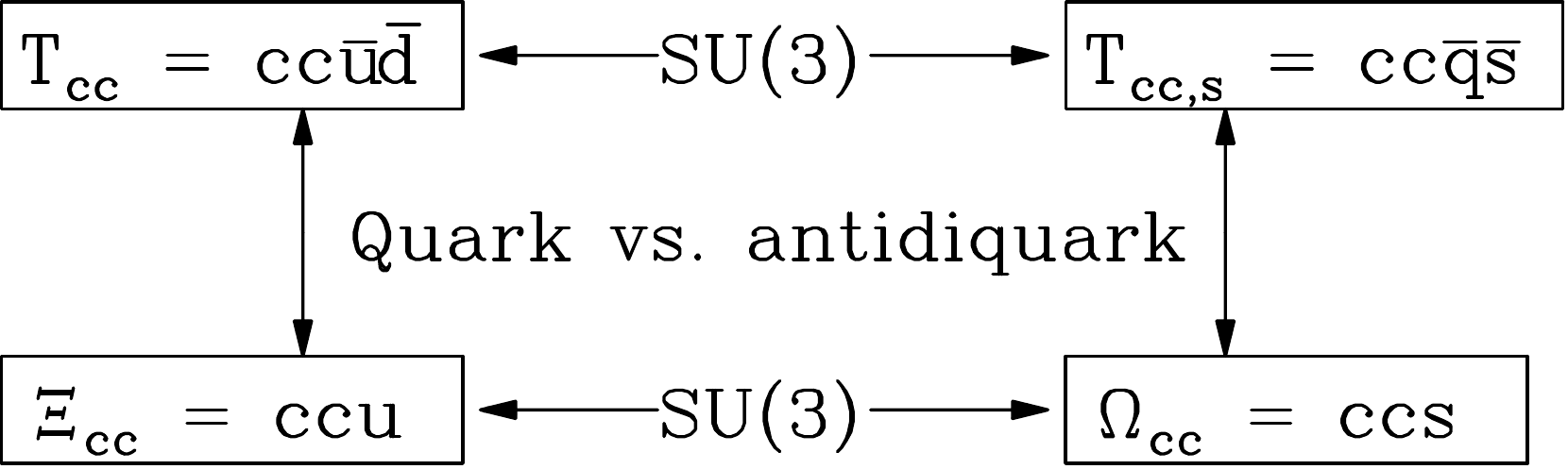}
\end{center}
\caption{Relations among processes involving production of states containing
two charmed quarks. Columns correspond to nonstrange (left) and strange (right)
particles.  Rows correspond to fragmentation of the $cc$ diquark into an
antidiquark (top) and a quark (bottom).  Processes in the left hand column
have been observed. \label{fig:ccfrag}}
\end{figure}

% This is Figure 4
\begin{figure}
\begin{center}
\includegraphics[width=0.48\textwidth]{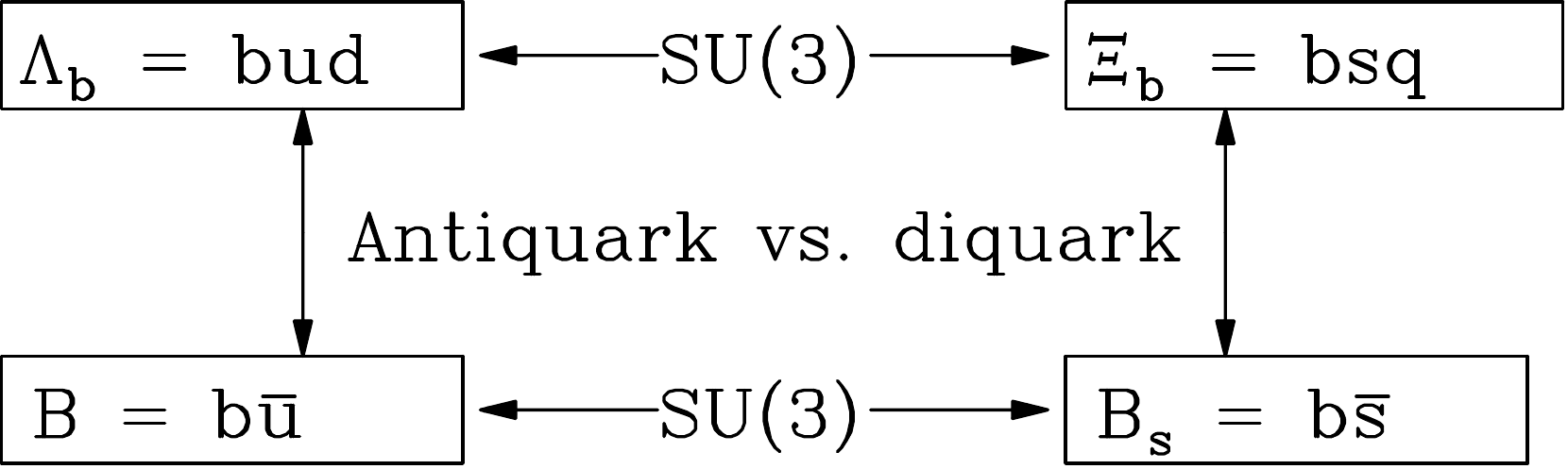}
\end{center}
\caption{Relations among processes involving production of states containing
one bottom quark. Columns correspond to nonstrange (left) and strange (right)
particles.  Rows correspond to fragmentation of the $b$ quark into a 
diquark (top) and an antiquark (bottom).  All processes have been observed.
\label{fig:bfrag}}
\end{figure}

% This is Figure 5
\begin{figure}
\begin{center}
	\includegraphics[width=0.48\textwidth]{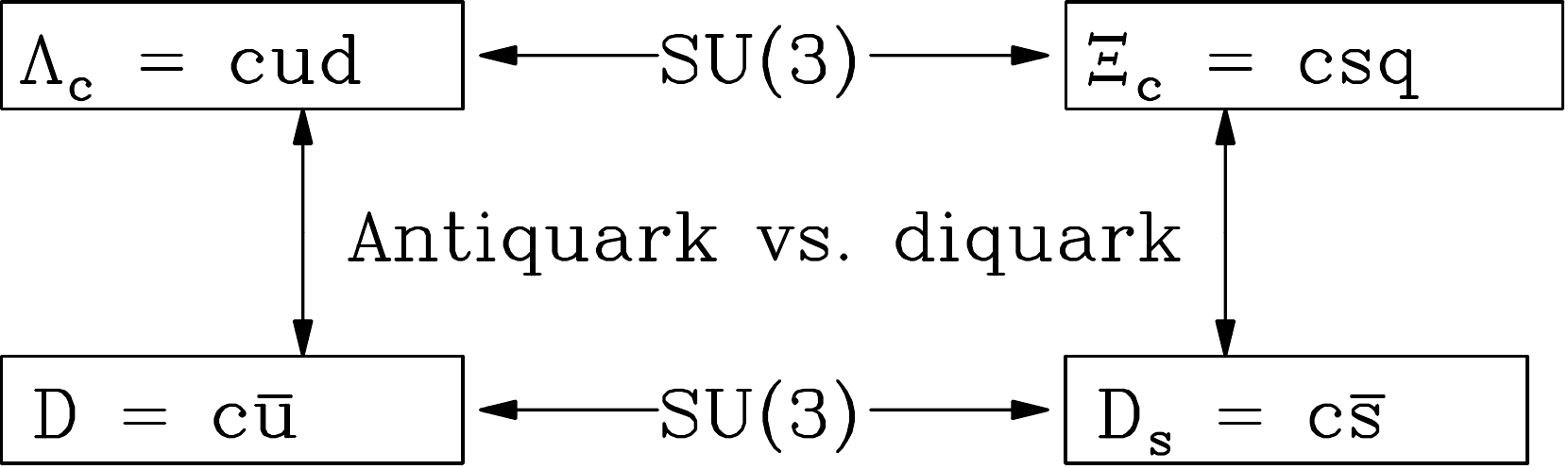}
\end{center}
\caption{Relations among processes involving production of states containing
one charmed quark. Columns correspond to nonstrange (left) and strange (right)
particles.  Rows correspond to fragmentation of the $c$ quark into a
diquark (top) and an antiquark (bottom).  All processes have been observed.
\label{fig:cfrag}}
\end{figure}

We gain some insight into the possible production rate of a $T_{cc,s}$ state
by comparison with $T_{cc}$ production.  This corresponds to the top $SU(3)$
relation in Fig.\ \ref{fig:ccfrag}. 
One can get a rough idea about the relevant relative fragmentation 
probabilities of a color antitriplet $cc$ diquark with mass $\sim 3.4$ GeV
by looking at the corresponding processes for a $b$ quark with mass 
$\sim 4.2$ GeV  \cite{ParticleDataGroup:2020ssz}.
These are described in Fig.\ \ref{fig:bfrag}; the corresponding relations
involving a charmed quark are shown for comparison in Fig.\ \ref{fig:cfrag}.
The fragmentation of a $b$ quark into
a strange quark accounts for roughly 1/8 of $b$ fragmentation into a nonstrange
quark \cite{CDF:2008yux,LHCb:2019fns,LHCb:2021qbv}.

In more detail:  For $b$ quark fragmentation at 13 TeV, Ref.\
\cite{LHCb:2019fns} quotes
\bea
f_s/(f_u+f_d) & = & 0.122 \pm 0.006~, \\
%JR
f_{\Lambda_b}/(f_u+f_d) & = & 0.259 \pm 0.018~,
\eea
%JR
where $f_q$ are the fragmentation functions of a $b$ quark to a $B_q$ hadron,
and similarly for the $\Lambda_b$.
The latter relation can be interpreted as describing the relative probability
of a heavy color (anti)triplet fragmenting into $[ud]$ (a ``good'' diquark)
vs.\ the sum of probabilities of fragmentation into $u$ or $d$ quarks.  So in
this case we need
\beq
f_{[ud]}/f_q = 2 \times 0.259 = 0.518 ~(q = u,~d)~.
\eeq
Further, from $f_s/(f_u + f_d)$ we have
\beq
f_s/f_q = 2 \times 0.122 = 0.244~.
\eeq
We then assume $SU(3)$ breaking  is the same in quarks and diquarks:
\beq
%JR       |
f_{qs}/f_{[ud]} = f_s/f_q = 0.244~.
\eeq
\beq
f_{qs}/f_q = f_{ud}/f_q \times f_{qs}/f_{ud} = 0.518\times0.244 = 0.126~.
\eeq
%JR ||| etc
The estimate of diquark-quark symmetry could be verified by comparing the
production of $T_{cc}$ with that of $\Xi_{cc}^{++}$, the former being 0.518
of the latter. LHCb can already provide an estimate for this ratio.

One must then take account of
differences in branching fractions into observable final states: $K^- \pi^+$
and $K^- \pi^+ \pi^- \pi^+$ (about 10\%) for $D^0$ and $K^+ K^- \pi^+$ (about
5\%) for $D_s$.  All told, one might expect the signal for the lightest
strange doubly charmed tetraquark to be about 1/16 of that for the $T_{cc}$.
Background might be higher because the $D_s$ signal is weaker
in the $T_{cc,s}$ than the second $D^0$ signal in the $T_{cc}$.

Another estimate of $T_{cc,s}$ production compares the fragmentation of $cc$
into $\bar q \bar s$ with $cc$ fragmentation into $u$, which gives rise to
the observed $\Xi_{cc}^{++}$ \cite{LHCb:2017iph}.  This comparison can involve
measurement of the intermediate fragmentation of $cc$
into $s$ and hence observation of the $\Omega_{cc} = ccs$.  We have noted
prospects for detecting the $\Omega_{cc}$ in Ref.\ \cite{Karliner:2014gca}.
As in the preceding discussion, a rough estimate of the relevant 
relative fragmentation probabilities can be obtained by comparing the
fragmentation of a $b$ quark into $\,\Xi_b\, (= bqs)$ and to a $B^-$ meson
$(=b\bar u)$. 

The observation by LHCb of a tetraquark $T_{cc}$ implies a rich spectrum of its
strange counterpart with quark content $c c \bar q \bar s$.  If the observed
$D^{*+} D^0$ enhancement is merely due to kinematics, one might expect the
prediction of $M(T_{cc,s}) = 4106$ MeV to fail.  A corresponding effect in
$M(D^0 D^0 K^+)$ around threshold would occur at $M(D_s(2714) D^0) = 4588$ MeV. 
If, on the other hand, one generalizes the approach of Ref.\
\cite{Karliner:2017qjm} to systems with quark content $c c \bar q \bar s$, one
expects several states decaying to a pair of charmed mesons, one nonstrange and
one strange, including one around 4100 MeV decaying to $D^* D_s$ and $D D_s^*$
and one not far from threshold decaying to $D^0 D^0 K^+$.
\medskip

\section*{ACKNOWLEDGMENTS}
We thank Vanya Belyaev and Tomasz Skwarnicki for correspondence on LHCb
$b$-quark fragmentation measurements, and an anonymous referee for
constructive comments..
The research of M.K. was supported in part by NSFC-ISF grant No.\ 3423/19.

\bibliography{tccs5}

\begin{thebibliography}{12}%
\makeatletter
\providecommand \@ifxundefined [1]{%
 \@ifx{#1\undefined}
}%
\providecommand \@ifnum [1]{%
 \ifnum #1\expandafter \@firstoftwo
 \else \expandafter \@secondoftwo
 \fi
}%
\providecommand \@ifx [1]{%
 \ifx #1\expandafter \@firstoftwo
 \else \expandafter \@secondoftwo
 \fi
}%
\providecommand \natexlab [1]{#1}%
\providecommand \enquote  [1]{``#1''}%
\providecommand \bibnamefont  [1]{#1}%
\providecommand \bibfnamefont [1]{#1}%
\providecommand \citenamefont [1]{#1}%
\providecommand \href@noop [0]{\@secondoftwo}%
\providecommand \href [0]{\begingroup \@sanitize@url \@href}%
\providecommand \@href[1]{\@@startlink{#1}\@@href}%
\providecommand \@@href[1]{\endgroup#1\@@endlink}%
\providecommand \@sanitize@url [0]{\catcode `\\12\catcode `\$12\catcode
  `\&12\catcode `\#12\catcode `\^12\catcode `\_12\catcode `\%12\relax}%
\providecommand \@@startlink[1]{}%
\providecommand \@@endlink[0]{}%
\providecommand \url  [0]{\begingroup\@sanitize@url \@url }%
\providecommand \@url [1]{\endgroup\@href {#1}{\urlprefix }}%
\providecommand \urlprefix  [0]{URL }%
\providecommand \Eprint [0]{\href }%
\providecommand \doibase [0]{http://dx.doi.org/}%
\providecommand \selectlanguage [0]{\@gobble}%
\providecommand \bibinfo  [0]{\@secondoftwo}%
\providecommand \bibfield  [0]{\@secondoftwo}%
\providecommand \translation [1]{[#1]}%
\providecommand \BibitemOpen [0]{}%
\providecommand \bibitemStop [0]{}%
\providecommand \bibitemNoStop [0]{.\EOS\space}%
\providecommand \EOS [0]{\spacefactor3000\relax}%
\providecommand \BibitemShut  [1]{\csname bibitem#1\endcsname}%
\let\auto@bib@innerbib\@empty
%</preamble>
\bibitem [{\citenamefont {Aaij}\ \emph
  {et~al.}(2021{\natexlab{a}})\citenamefont {Aaij} \emph
  {et~al.}}]{LHCb:2021vvq}%
  \BibitemOpen
  \bibfield  {author} {\bibinfo {author} {\bibfnamefont {Roel}\ \bibnamefont
  {Aaij}} \emph {et~al.} (\bibinfo {collaboration} {LHCb}),\ }\bibfield
  {title} {\enquote {\bibinfo {title} {{Observation of an exotic narrow doubly
  charmed tetraquark}},}\ }\href@noop {} {\  (\bibinfo {year}
  {2021}{\natexlab{a}})},\ \Eprint {http://arxiv.org/abs/2109.01038}
  {arXiv:2109.01038 [hep-ex]} \BibitemShut {NoStop}%
\bibitem [{\citenamefont {Aaij}\ \emph
  {et~al.}(2021{\natexlab{b}})\citenamefont {Aaij} \emph
  {et~al.}}]{LHCb:2021auc}%
  \BibitemOpen
  \bibfield  {author} {\bibinfo {author} {\bibfnamefont {Roel}\ \bibnamefont
  {Aaij}} \emph {et~al.} (\bibinfo {collaboration} {LHCb}),\ }\bibfield
  {title} {\enquote {\bibinfo {title} {{Study of the doubly charmed tetraquark
  $T_{cc}^+$}},}\ }\href@noop {} {\  (\bibinfo {year} {2021}{\natexlab{b}})},\
  \Eprint {http://arxiv.org/abs/2109.01056} {arXiv:2109.01056 [hep-ex]}
  \BibitemShut {NoStop}%
\bibitem [{\citenamefont {Karliner}\ and\ \citenamefont
  {Rosner}(2017)}]{Karliner:2017qjm}%
  \BibitemOpen
  \bibfield  {author} {\bibinfo {author} {\bibfnamefont {Marek}\ \bibnamefont
  {Karliner}}\ and\ \bibinfo {author} {\bibfnamefont {Jonathan~L.}\
  \bibnamefont {Rosner}},\ }\bibfield  {title} {\enquote {\bibinfo {title}
  {{Discovery of doubly-charmed $\Xi_{cc}$ baryon implies a stable ($b b
  \bar{u} \bar{d}$) tetraquark}},}\ }\href {\doibase
  10.1103/PhysRevLett.119.202001} {\bibfield  {journal} {\bibinfo  {journal}
  {Phys. Rev. Lett.}\ }\textbf {\bibinfo {volume} {119}},\ \bibinfo {pages}
  {202001} (\bibinfo {year} {2017})},\ \Eprint
  {http://arxiv.org/abs/1707.07666} {arXiv:1707.07666 [hep-ph]} \BibitemShut
  {NoStop}%
\bibitem [{\citenamefont {Karliner}\ and\ \citenamefont
  {Rosner}(2014)}]{Karliner:2014gca}%
  \BibitemOpen
  \bibfield  {author} {\bibinfo {author} {\bibfnamefont {Marek}\ \bibnamefont
  {Karliner}}\ and\ \bibinfo {author} {\bibfnamefont {Jonathan~L.}\
  \bibnamefont {Rosner}},\ }\bibfield  {title} {\enquote {\bibinfo {title}
  {{Baryons with two heavy quarks: Masses, production, decays, and
  detection}},}\ }\href {\doibase 10.1103/PhysRevD.90.094007} {\bibfield
  {journal} {\bibinfo  {journal} {Phys. Rev. D}\ }\textbf {\bibinfo {volume}
  {90}},\ \bibinfo {pages} {094007} (\bibinfo {year} {2014})},\ \Eprint
  {http://arxiv.org/abs/1408.5877} {arXiv:1408.5877 [hep-ph]} \BibitemShut
  {NoStop}%
\bibitem [{\citenamefont {Aaij}\ \emph {et~al.}(2017)\citenamefont {Aaij} \emph
  {et~al.}}]{LHCb:2017iph}%
  \BibitemOpen
  \bibfield  {author} {\bibinfo {author} {\bibfnamefont {Roel}\ \bibnamefont
  {Aaij}} \emph {et~al.} (\bibinfo {collaboration} {LHCb}),\ }\bibfield
  {title} {\enquote {\bibinfo {title} {{Observation of the doubly charmed
  baryon $\Xi_{cc}^{++}$}},}\ }\href {\doibase 10.1103/PhysRevLett.119.112001}
  {\bibfield  {journal} {\bibinfo  {journal} {Phys. Rev. Lett.}\ }\textbf
  {\bibinfo {volume} {119}},\ \bibinfo {pages} {112001} (\bibinfo {year}
  {2017})},\ \Eprint {http://arxiv.org/abs/1707.01621} {arXiv:1707.01621
  [hep-ex]} \BibitemShut {NoStop}%
\bibitem [{\citenamefont {Francis}\ \emph {et~al.}(2021)\citenamefont
  {Francis}, \citenamefont {de~Forcrand}, \citenamefont {Lewis},\ and\
  \citenamefont {Maltman}}]{Francis:2021vrr}%
  \BibitemOpen
  \bibfield  {author} {\bibinfo {author} {\bibfnamefont {Anthony}\ \bibnamefont
  {Francis}}, \bibinfo {author} {\bibfnamefont {Philippe}\ \bibnamefont
  {de~Forcrand}}, \bibinfo {author} {\bibfnamefont {Randy}\ \bibnamefont
  {Lewis}}, \ and\ \bibinfo {author} {\bibfnamefont {Kim}\ \bibnamefont
  {Maltman}},\ }\bibfield  {title} {\enquote {\bibinfo {title} {{Diquark
  properties from full QCD lattice simulations}},}\ }\href@noop {} {\
  (\bibinfo {year} {2021})},\ \Eprint {http://arxiv.org/abs/2106.09080}
  {arXiv:2106.09080 [hep-lat]} \BibitemShut {NoStop}%
\bibitem [{\citenamefont {Zyla}\ \emph {et~al.}(2020)\citenamefont {Zyla} \emph
  {et~al.}}]{ParticleDataGroup:2020ssz}%
  \BibitemOpen
  \bibfield  {author} {\bibinfo {author} {\bibfnamefont {P.~A.}\ \bibnamefont
  {Zyla}} \emph {et~al.} (\bibinfo {collaboration} {Particle Data Group}),\
  }\bibfield  {title} {\enquote {\bibinfo {title} {{Review of Particle
  Physics}},}\ }\href {\doibase 10.1093/ptep/ptaa104} {\bibfield  {journal}
  {\bibinfo  {journal} {PTEP}\ }\textbf {\bibinfo {volume} {2020}},\ \bibinfo
  {pages} {083C01} (\bibinfo {year} {2020})}\BibitemShut {NoStop}%
\bibitem [{\citenamefont {Ren}\ \emph {et~al.}(2021)\citenamefont {Ren},
  \citenamefont {Wu},\ and\ \citenamefont {Zhu}}]{ren:2021dsi}%
  \BibitemOpen
  \bibfield  {author} {\bibinfo {author} {\bibfnamefont {Huimin}\ \bibnamefont
  {Ren}}, \bibinfo {author} {\bibfnamefont {Fan}\ \bibnamefont {Wu}}, \ and\
  \bibinfo {author} {\bibfnamefont {Ruilin}\ \bibnamefont {Zhu}},\ }\bibfield
  {title} {\enquote {\bibinfo {title} {{Hadronic molecule interpretation of
  $T^+_{cc}$ and its beauty-partners}},}\ }\href@noop {} {\  (\bibinfo {year}
  {2021})},\ \Eprint {http://arxiv.org/abs/2109.02531} {arXiv:2109.02531
  [hep-ph]} \BibitemShut {NoStop}%
\bibitem [{\citenamefont {Eichten}\ and\ \citenamefont
  {Quigg}(2017)}]{Eichten:2017ffp}%
  \BibitemOpen
  \bibfield  {author} {\bibinfo {author} {\bibfnamefont {Estia~J.}\
  \bibnamefont {Eichten}}\ and\ \bibinfo {author} {\bibfnamefont {Chris}\
  \bibnamefont {Quigg}},\ }\bibfield  {title} {\enquote {\bibinfo {title}
  {{Heavy-quark symmetry implies stable heavy tetraquark mesons $Q_iQ_j \bar
  q_k \bar q_l$}},}\ }\href {\doibase 10.1103/PhysRevLett.119.202002}
  {\bibfield  {journal} {\bibinfo  {journal} {Phys. Rev. Lett.}\ }\textbf
  {\bibinfo {volume} {119}},\ \bibinfo {pages} {202002} (\bibinfo {year}
  {2017})},\ \Eprint {http://arxiv.org/abs/1707.09575} {arXiv:1707.09575
  [hep-ph]} \BibitemShut {NoStop}%
\bibitem [{\citenamefont {Aaltonen}\ \emph {et~al.}(2008)\citenamefont
  {Aaltonen} \emph {et~al.}}]{CDF:2008yux}%
  \BibitemOpen
  \bibfield  {author} {\bibinfo {author} {\bibfnamefont {T.}~\bibnamefont
  {Aaltonen}} \emph {et~al.} (\bibinfo {collaboration} {CDF}),\ }\bibfield
  {title} {\enquote {\bibinfo {title} {{Measurement of Ratios of Fragmentation
  Fractions for Bottom Hadrons in $p \bar{p}$ Collisions at $\sqrt{s}$ =
  1.96-TeV}},}\ }\href {\doibase 10.1103/PhysRevD.77.072003} {\bibfield
  {journal} {\bibinfo  {journal} {Phys. Rev. D}\ }\textbf {\bibinfo {volume}
  {77}},\ \bibinfo {pages} {072003} (\bibinfo {year} {2008})},\ \Eprint
  {http://arxiv.org/abs/0801.4375} {arXiv:0801.4375 [hep-ex]} \BibitemShut
  {NoStop}%
\bibitem [{\citenamefont {Aaij}\ \emph {et~al.}(2019)\citenamefont {Aaij} \emph
  {et~al.}}]{LHCb:2019fns}%
  \BibitemOpen
  \bibfield  {author} {\bibinfo {author} {\bibfnamefont {Roel}\ \bibnamefont
  {Aaij}} \emph {et~al.} (\bibinfo {collaboration} {LHCb}),\ }\bibfield
  {title} {\enquote {\bibinfo {title} {{Measurement of $b$ hadron fractions in
  13 TeV $pp$ collisions}},}\ }\href {\doibase 10.1103/PhysRevD.100.031102}
  {\bibfield  {journal} {\bibinfo  {journal} {Phys. Rev. D}\ }\textbf {\bibinfo
  {volume} {100}},\ \bibinfo {pages} {031102} (\bibinfo {year} {2019})},\
  \Eprint {http://arxiv.org/abs/1902.06794} {arXiv:1902.06794 [hep-ex]}
  \BibitemShut {NoStop}%
\bibitem [{\citenamefont {Aaij}\ \emph
  {et~al.}(2021{\natexlab{c}})\citenamefont {Aaij} \emph
  {et~al.}}]{LHCb:2021qbv}%
  \BibitemOpen
  \bibfield  {author} {\bibinfo {author} {\bibfnamefont {Roel}\ \bibnamefont
  {Aaij}} \emph {et~al.} (\bibinfo {collaboration} {LHCb}),\ }\bibfield
  {title} {\enquote {\bibinfo {title} {{Precise measurement of the~$f_s/f_d$
  ratio of fragmentation fractions and of $B^0_s$ decay branching
  fractions}},}\ }\href {\doibase 10.1103/PhysRevD.104.032005} {\bibfield
  {journal} {\bibinfo  {journal} {Phys. Rev. D}\ }\textbf {\bibinfo {volume}
  {104}},\ \bibinfo {pages} {032005} (\bibinfo {year} {2021}{\natexlab{c}})},\
  \Eprint {http://arxiv.org/abs/2103.06810} {arXiv:2103.06810 [hep-ex]}
  \BibitemShut {NoStop}%
\end{thebibliography}%
\end{document}